\documentclass{article}
\pdfoutput=1
\title{Solar Influence on the North Atlantic Oscillation - Initial Results}
\author{Sally Dacie}
\usepackage{graphicx}
\usepackage{gensymb}
\usepackage{fixltx2e}
\usepackage{amsmath}
\usepackage{natbib}
\usepackage[margin=1in]{geometry}

\def\eg{{\it e.g., }}
\def\ie{{\it i.e.}}

\begin{document}
\maketitle

\section*{Abstract}

Some initial investigations into various atmospheric phenomena and the influence of the solar cycle on weather have been made. Strongly negative North Atlantic Oscillation (NAO) indices, which cause cold and dry winter weather in North West Europe, rarely occur during periods of high solar activity. Coupling between the troposphere and stratosphere is discussed, particularly in the context of Polar-night jet oscillation events \citep[defined by][]{Hitchcock13} and the Quasi-Biennial Oscillation. The energy of North Atlantic hurricanes (as indicated by the Accumulated Cyclone Energy Index, ACE) is also linked to solar activity, via UV heating at the tropopause \citep{Elsner10}, and is suggested as a possible mechanism through which solar activity could influence the NAO. Finally the lack of solar influence on the NAO before $\sim$ 1950 is addressed, with a possible cause being the smaller solar cycle amplitudes. This short report contains several ideas, which may be worth pursuing further.

\section{Introduction to the North Atlantic Oscillation}

The North Atlantic Oscillation (NAO) is measured as the difference in pressure between the Icelandic Low and the Bermuda-Azores high. Positive NAO occurs with a large pressure difference and a strong storm track, which brings wet and stormy weather to North West Europe; Negative NAO has a small pressure difference and is associated with dry weather in North West Europe. The NAO is associated with the Artic Oscillation (AO), which is defined as the leading Empirical Orthogonal Function of the NAO and extends up into the stratosphere. Many researches consider the NAO as a ``bell", because it is an amplified response to small forcings and is amplified beyond what one might expect compared to the stratospheric / upper tropospheric signal (Baldwin, at the Stratosphere-Troposphere Coupling Meeting of the Royal Meteorological Society 2015). 

Two factors, which force the NAO, are the solar cycle and to a lesser extent the Quasi-Biennial Oscillation (QBO); the oscillation between easterly and westerly winds in the equatorial stratosphere. Solar activity is found to influence the NAO such that strongly negative winter NAO values (which cause cold and dry conditions in North West Europe) rarely occur during periods of high solar activity.

\section{Stratosphere-Troposphere Coupling}

The direction of influence - whether the troposphere has greater influence on the stratosphere than the stratosphere on the troposphere - is fundamental to our understanding of atmospheric behaviour and predictability. Previously, focus was on the upward influence of the troposphere, in particular the upward propagation of tropospheric planetary waves \citep[\eg][]{McIntyre83}. More recently, studies \citep[\eg][]{Boer08} have focused on the downward influence of the stratosphere with the aim of using stratospheric phenomena (such as the Quasi-Biennial Oscillation) to improve predictability of tropospheric weather patterns. As the word `coupling' suggests, the link between the stratosphere and the troposphere goes in both directions.

\subsection{Polar-Night Jet Oscillation Events}

Sudden Stratospheric Warmings (SSWs) are a clear example of stratosphere-troposphere coupling. They consist of a complete break down of the main vortex. \citet{Hitchcock13} defined Polar-Night Jet Oscillation events (PJO events) as extended-time-scale recoveries from stratospheric sudden warmings. Due to their relatively long time scales and radiative dynamics, their influence on the troposphere is greater than that of weak vortex events.

\begin{figure}
	\includegraphics[scale=0.5]{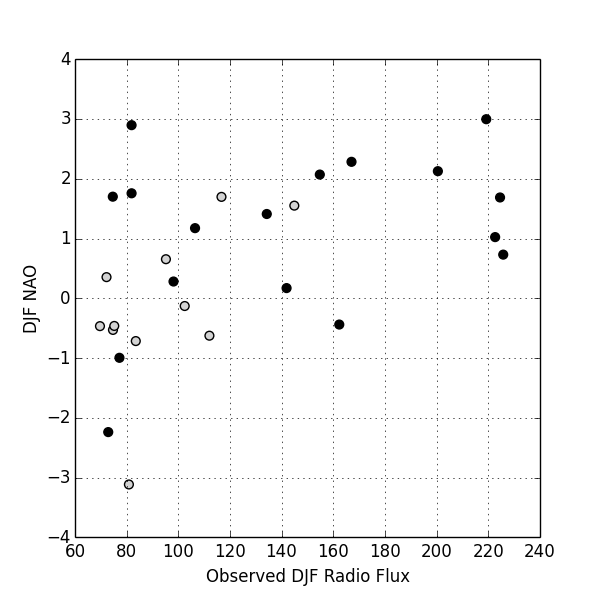}
	\caption{DJF NAO is plotted against observed DJF Radio flux, for the winters 1982/83 to 2009/10. Winters (November to March) in which a PJO event occurs are marked with gray points, and other winters are black. NAO data was taken from the Climate Research Unit of the University of East Anglia (UEA) and Radio flux data were the Penticton values from NOAA (the National Oceanic and Atmospheric Administration).}
	\label{PJO}
\end{figure}

I have found that the occurrence of PJO events, linked to a weakened polar vortex, are much more likely under low solar activity. Figure \ref{PJO} shows this link and suggests that PJO events are more strongly linked to low solar activity than to negative winter NAO; that solar activity is influencing the stratosphere independently of the NAO (rather than due to upward propagating planetary waves caused as a result of a ``solar-induced'' NAO). SSWs and particularly PJO events have a strong influence on lower tropospheric conditions \citep{Fereday12,Hitchcock13}, and could be causing an NAO signal. From Figure \ref{PJO}, however, due to the limited data available (only 11 years with PJO events) and inherent variability, it is hard to draw conclusions on the direction of influence. Looking at the timing of these PJO events with respect to changes in the NAO on timescales of weeks (rather than looking just at the winter average) may provide some insight here.

Many researches consider the Quasi-Biennial Oscillation (QBO) to exert an influence on the polar vortex; during easterly QBO the vortex is more disturbed \citep[\eg][]{Boer08,Marshall09} and the probability of SSWs is increased \citep[\eg][]{Dameris90,Fereday12}. The El Ni\~no Southern Oscillation (ENSO) is also thought to influence the polar vortex \citep[\eg][]{Free09,Butler11}. Despite this, my initial investigations suggest that solar activity has a much stronger influence on PJO events than either the QBO or ENSO. 

\subsection{Stratospheric Winds}

The Quasi-Biennial Oscillation (QBO) is considered to be driven by the transfer of momentum from waves propagating upward from the troposphere. This is suggestive of a strong upward influence of the troposphere on the stratosphere. Studies show that there is no (or very little) solar cycle modulation of QBO wind amplitudes, but that the duration of the QBO phases are longer under low solar activity than high \citep{Soukharev01,Pascoe05}. I speculate that this could be an indirect effect of solar activity via the NAO response, as a positive NAO (associated with high solar activity) results in an increase in equatorial refraction of upward-propagating planetary waves \citep{Ambaum02}.

Although there is no solar modulation of the QBO wind strengths, some studies have found a subtropical stratospheric wind response to the solar cycle. \citet{Crooks05}, using ERA-40 reanalysis, found a significant subtropical zonal wind response of up to 6 ms\textsuperscript{-1} in the upper stratosphere, and \citet{Kodera02} found a solar cycle influence on the stratopause subtropical jet, with speeds increasing from 51 to 59 ms\textsuperscript{-1} in the Northern Hemisphere. Lower down in the atmosphere, Scaife (at the Stratosphere-Troposphere Coupling Meeting of the Royal Meteorological Society 2015) proposed that an acceleration of the subtropical jet (which is located just below the tropopause) may be responsible for causing the NAO. However, it is unclear to me, whether a strong subtropical jet results in or results from the NAO.

\section{Time Lag Investigations}

\begin{figure}
	\includegraphics[scale=0.4]{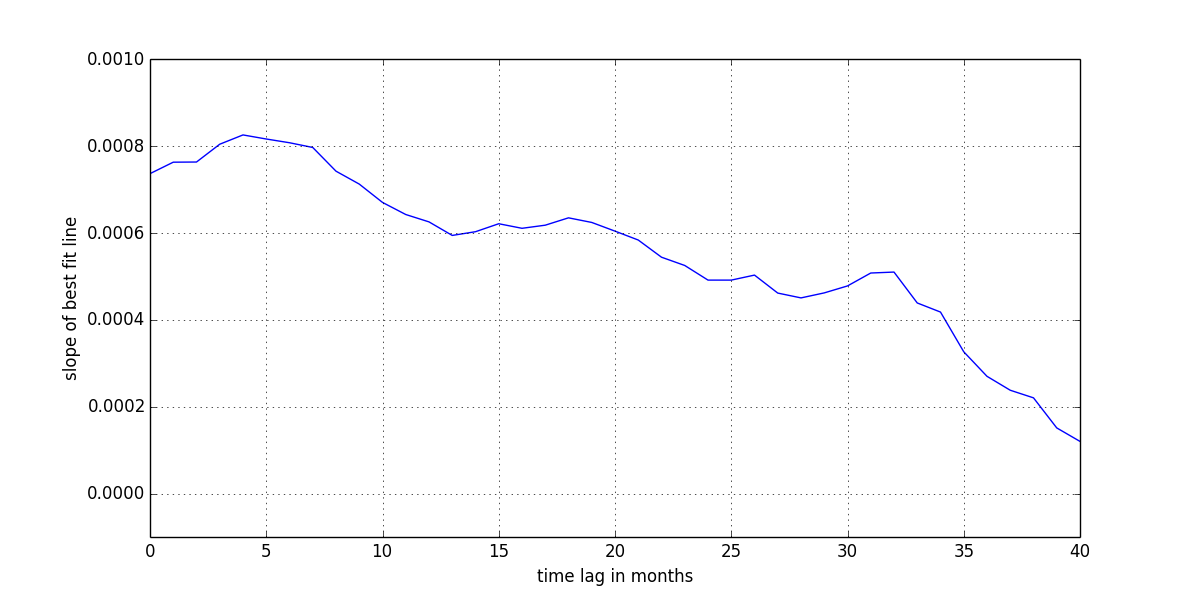}
	\caption{A straight line of best fit (using the least squares method) was plotted for graphs of DJF NAO index against 5 month radio flux averages at different lag times, with for example radio flux from November to March corresponding to zero lag, and radio flux from October to February corresponding to a one month lag. The gradients of these best fit lines were then plotted against time lag, to produce the figure shown. DJF NAO values from 1950/51 to 2013/14 were used. NAO data was taken from the Climate Research Unit at the University of East Anglia. 10.7 cm Radio flux data was from NOAA.}
	\label{timelagslope}
\end{figure}

In order to better understand a mechanism for solar influence on the NAO, initial investigations into the optimal time lag between solar activity and DJF NAO response were made. A peak in the strength of the relationship was found at a time lag of around 4 to 5 months, see Figure \ref{timelagslope}. Other studies \citep{Qun93,Scaife13} have found a longer lagged response of the Atlantic subtropical high and the NAO to solar variability of 2 to 4 years and some \citep[\eg][]{Scaife13} suggest that this is related to inertia in the system caused by slowly warming sea temperatures as a result of higher solar activity. \citet{White97} found a phase lag of upper ocean temperature of 30 to 50 degrees (equivalent to $\sim$ 1 to 2 years) and up to 75 degrees ( $\sim$ 3 years) at depths of 80-160 m. I think ocean heating is unlikely to be the main cause of the NAO response, because SST response to the solar cycle is small, especially compared to changes caused by global warming, which, over the period 1971 to 2010, has lead to an increase of 0.11\degree K per decade in the globally averaged ocean's upper 75 m \citep{IPCC13}. Although Figure \ref{timelagslope} still shows a NAO response out to $\sim$ 3 years, the relationship is found to weaken after $\sim$ 6 months, and it is possible that this extended response is due to autocorrelation of the NAO. 

On a decadal time scale, solar signals only penetrate the ocean to a depth of 80-160 m \citep{White97}. Perhaps on longer timescales the deep ocean also feels the affects of reduced irradiance. This could explain the larger changes in climate associated with solar variability on longer time scales, such as during the Maunder Minimum.

\section{Solar Influence on Wind}

An observational study \citep{Elsner10} investigated a link between hurricane activity over the Caribbean and solar activity on a daily timescale. The authors proposed that the apparent decrease in hurricane intensity during periods of high solar activity could result from UV heating near the top of the cyclone reducing the convective available potential energy (CAPE). They also suggested that tropical cyclones could act to amplify small changes in solar radiation. A further study \citep{Hodges14} found a regional response of hurricane frequency to the solar cycle.

\begin{figure}
	\includegraphics[scale=0.5]{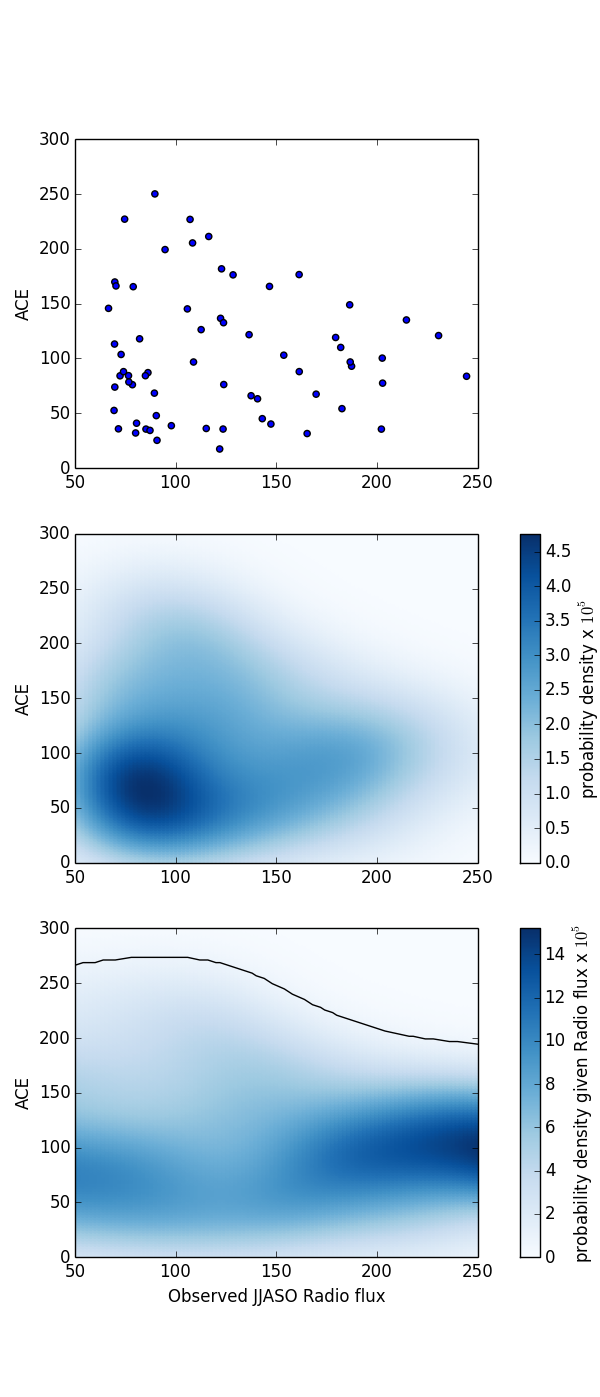}
	\caption{The top plot in this figure shows ACE index against observed JJASO Radio flux for the years 1950 to 2014. To make the middle figure, a smoothing (a Gaussian independently in both x and y) and a scaling has been applied to each data point to create a physically realistic smooth probability distribution map. In the bottom figure, these `probabilities' have been divided by the `probability' of obtaining the corresponding x value (calculated using the same data points and with the same smoothing applied in the x direction), such that the `probability density given a certain value of radio flux' is shown. The black line on the bottom figure is a contour, joining points of `probability density' between 0.8 and 0.85 x 10\textsuperscript{5}. Radio flux data was taken from NOAA and the ACE used was the same as that used by \citet{Boyd14} with an additional value provided by Saunders (personal communication) for 2014.}
	\label{sandpiles}
\end{figure}

The Accumulated Cyclone Energy Index (ACE) gives an indication of the cumulative energy of all North Atlantic hurricanes during a given season (June to November). The top plot in Figure \ref{sandpiles} shows that none of the highest ACE seasons have occurred during periods of high solar activity. Although the mean ACE value is virtually independent of solar activity, the bottom plot of Figure \ref{sandpiles} 
\footnote{Something that could be improved on is the production of the `probability' plots in Figure \ref{sandpiles}. While smoothing of the data points seems like a reasonable idea to get an idea of the overall distribution, a more sophisticated technique would apply different smoothing to the points depending on the density of the region they were in, such that areas with a higher density of points would retain more information in the probability distribution, where as points in low density regions would be smoothed more to fill in for `data gaps'.} 
shows that given a high JJASO Radio flux value, it is highly likely that the ACE value for the season will be between $\sim$ 50 and $\sim$ 150, whereas for lower JJASO Radio flux values the variance in ACE is much larger. Seasons of extreme hurricane activity tend to occur under low solar activity. Plotting hurricane number against sunspot number, which allowed an extension back to 1878, produced a distribution similar to that in the top plot of Figure \ref{sandpiles}, as did a plot of ACE / hurricane number against sunspot number. Both frequency and strength of hurricanes are found to be affected by solar activity.

Hurricane activity is known to be strongly correlated to sea surface temperature (SST). Although sea surface temperature is affected by solar activity, there is a system inertia causing a time lag between solar activity and SST, thought to be 1 - 3 years \citep{White97}. \citet{Elsner99} found that solar activity explains some of the variability of baroclinically enhanced hurricanes and speculated that this could be due to an increase in evaporation at the ocean surface under high solar activity. If such a mechanism were influential, one would expect to see an increase in hurricane activity at a lag of $\sim$ 1 - 3 years with respect to solar activity. This study finds a more immediate response. The other factor which suggests that a solar modulation of hurricane activity is not predominantly caused by anomalous SSTs is that solar activity only causes relatively small changes in SSTs, compared to the warming observed between $\sim$1910 - 1940 and $\sim$1970 onwards. Adjusting ACE for the global averaged temperature, according to a linear regression model based on data from 1950 to 2014, produced little change in the `probability' plots shown in Figure \ref{sandpiles}. Perhaps a change would be seen if this adjustment were instead made for North Atlantic SST, however I expect the plots would remain similar.

\begin{figure}
	\includegraphics[scale=0.5]{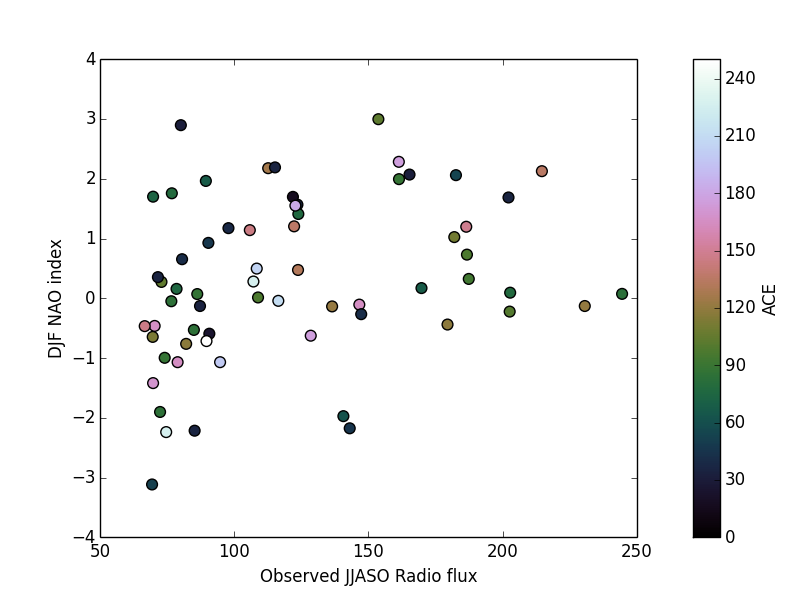}
	\caption{DJF NAO index was plotted against observed JJASO Radio flux with the colours of the points being determined by the ACE index (of the hurricane season preceding the winter). Radio flux and ACE indices from 1950 to 2013 and the corresponding NAO winters from 1950/51 to 2013/14 were used. NAO data was from the Climate Research Unit at UEA, Radio flux data from NOAA and the ACE used was the same as that used by \citet{Boyd14}.}	
	\label{withACE}
\end{figure}

\citet{Boyd14}, undertaking a statistical study, found an inverse relationship between the strength of the North Atlantic hurricane season (as measured by ACE) and the following winter NAO. \citet{Czaja02} investigated the covariances between monthly SST and 500 hPa height anomalies with time lag. While the dominant signal was atmospheric forcing of SST anomalies (corresponding to NAO influence on hurricanes), they also found statistically significant covariances with a North Atlantic Horseshoe pattern of SSTs leading the NAO by several months \citep{Czaja02}. Work by \citet{Gastineau13} supports this, and the authors state that the horseshoe pattern of SSTs is partially caused by internal variability of the atmosphere. This internal variability would include hurricanes and other small scale weather systems, so it is possible that solar activity is affecting wind speeds which in turn affect SST patterns, in such a way (\ie, by forming the North Atlantic Horseshoe) as to influence the upcoming winter NAO. Figure \ref{withACE} shows both that years of high ACE index (pale coloured points) occur during low solar activity and that the high ACE index acts to lower the NAO index. Years of reasonably high ACE index ($\sim$ 135 - 195) occurring during middle to high solar flux summers could result from the smaller effect of increased SSTs. These years are also ones of more positive NAO, in general, and warmer SSTs have been linked to more positive NAO winters in the context of global warming \citep{Hoerling01,Rind05}.

A similar process could also be happening to cause the slight solar signal in the El Ni\~no Southern Oscillation indices, Oceanic Ni\~no Index (ONI) and Southern Oscillation Index (SOI).

Interestingly, the QBO is also thought to influence tropospheric winds and vorticity \citep{Collimore03,Huesmann01}, and \citet{Ho09} even showed an influence on tropical cyclones, but whether this could be a viable mechanism for QBO influence on the NAO has not been studied.

\section{Lack of Solar Signal Before $\sim$1950}

Perhaps disappointingly, a lack of solar influence on the NAO before $\sim$ 1950 is found. This is in agreement with the study by \citet{Rodwell03} and is illustrated in Figure 2 of his paper. This lack of signal cannot be explained by poor sunspot data quality before that time \citep[see][]{Clette14}. One possible explanation is to do with the size of the solar cycle amplitudes; that since solar cycle 18, which started in January 1944, the cycle maxima have been consistently high.

\begin{figure}
	\includegraphics[scale=0.7]{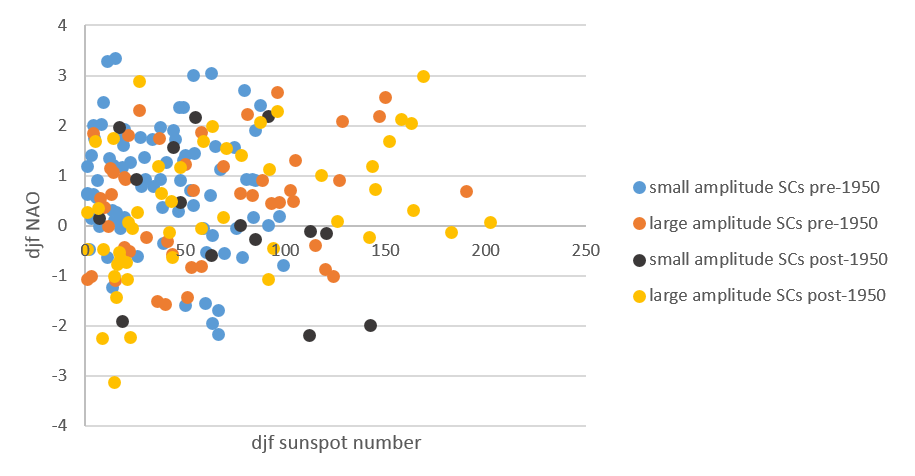}
	\caption{DJF NAO index was plotted against DJF Sunspot Number for winters 1823/24 to 2013/14. Different coloured points were used depending on solar cycle amplitude, both before and after 1949/50. Large amplitude cycles are defined as ones with Sunspot Number greater than 100 as in Figure 62 of `Revisiting the Sunspot Number' \citep{Clette14}. Cycles are defined to start and end 2 years later than the minimum (2 years later than the standard definition).  NAO data was taken from the Climate Research Unit at the UEA and Sunspot Number data from SILSO (Sunspot Index and Long-term Solar Observations).}	
	\label{NAOwithSCamplitude}
\end{figure}

As can be seen in Figure \ref{NAOwithSCamplitude}, for small amplitude cycles (which don't reach high sunspot values), the NAO signal vanishes. A reduced NAO signal would be expected, given that by definition the sunspot number does not reach a particularly high value during small amplitude cycles, and that the whole range of NAO index values can occur under low sunspot number (at least during this period 1823/24 to 2013/14). However, one would not necessary expect the NAO signal to vanish completely. Moreover, the occurrence of negative NAO values during solar minima is emphasised, when looking just at winters during large amplitude solar cycles. This plot also shows that, separate to the solar cycle amplitude effect, the strength of the solar-NAO link has increased post 1950. However the points belonging to the large amplitude solar cycles before this time still show an upward trend; the least-squares best fit line has a gradient of 0.0064 compared to 0.0091 for the post 1950 large amplitude cycles.

It is also possible that changes in the underlying climate regime could be affecting the presence of a solar signal. Such a change may have occurred during the beginning of the 20th century, when global temperatures increased due to a combination of greenhouse gas emissions and a large amount of internal variability \citep{Delworth00}. The more recent period of warming and the fact that heat is being stored deep in the oceans \citep{Chen14} has large implications for the climate and may cause a change in the relationship between solar activity and weather.

\section{Summary and Suggestions for Further Research}

In terms of the solar induced NAO signal, there are two main questions. The first of these addresses the mechanism: how does a solar signal manifest itself in the winter NAO, a mid to high latitude weather pattern? My initial results suggest that hurricanes and other localised wind systems, influenced by UV heating at the tropopause, could set up the sea surface temperature and pressure patterns associated with the NAO. One of the key arguments supporting this result is the optimal time lag of $\sim$ 5 - 6 months, but more detailed analysis into the time lag is needed. A problem associated with this is how to quantify the strength of the relationship between two variables using a scatter plot which is not particularly well suited to a best fit line (as the variance of points changes along the line: the whole range of NAO values occur under low solar activity, but only upper NAO values occur under high). A way of getting around this, could be to use pressure maps rather than a single index for the NAO. Then one would look at how the pressure patterns change with varying solar activity and a principle component could be identified, presumably reflective of the NAO, which could be used to determine the optimal time lag. This method would be similar to the maximum covariance analysis used by \citet{Czaja02}.

The second question is why the solar signal in the NAO disappears before $\sim$ 1950. Two possibilities have been suggested here; smaller amplitude solar cycles and a changing underlying climate. Perhaps when knowledge of the mechanism behind the solar-NAO link improves, this question will also be answered. Equally, an understanding of why the signal vanishes could provide clues as to the mechanism.

Something else that would potentially be interesting, would be an investigation into a upward influence of the NAO, for example on QBO downward propagation speeds and Polar-Night Jet Oscillation events. If the NAO was found to affect QBO phase lengths, it may be the cause of the link between solar activity and QBO phase propagation speeds, which was seen by \citet{Soukharev01} among others. Once again, it will be important to look at the time lag between tropospheric and stratospheric phenomena, in order to infer causality.

\section*{Acknowledgements}

I would like to thank Professors Mark Saunders and Lidia van Driel-Gesztelyi, without whom these initial investigations would never have been made.

\bibliographystyle{ametsoc}
\bibliography{MyCollection}

\begin{thebibliography}{32}
\providecommand{\natexlab}[1]{#1}
\providecommand{\url}[1]{\texttt{#1}}
\providecommand{\urlprefix}{URL }
\expandafter\ifx\csname urlstyle\endcsname\relax
  \providecommand{\doi}[1]{doi:\discretionary{}{}{}#1}\else
  \providecommand{\doi}{doi:\discretionary{}{}{}\begingroup
  \urlstyle{rm}\Url}\fi
\providecommand{\eprint}[2][]{\url{#2}}

\bibitem[{Ambaum and Hoskins(2002)}]{Ambaum02}
Ambaum, M. H.~P. and B.~J. Hoskins, 2002: {The NAO troposphere-stratosphere
  connection}. \textit{Journal of Climate}, \textbf{15}, 1969--1978,
  \doi{10.1175/1520-0442(2002)015<1969:TNTSC>2.0.CO;2}.

\bibitem[{Boer and Hamilton(2008)}]{Boer08}
Boer, G.~J. and K.~Hamilton, 2008: {QBO influence on extratropical predictive
  skill}. \textit{Climate Dynamics}, \textbf{31}, 987--1000,
  \doi{10.1007/s00382-008-0379-5},
  \urlprefix\url{http://link.springer.com/10.1007/s00382-008-0379-5}.

\bibitem[{Boyd(2014)}]{Boyd14}
Boyd, J.~A., 2014: {The anti-correlation between North Atlantic hurricane
  seasons and European windstorm seasons}. Tech. rep.

\bibitem[{Butler and Polvani(2011)}]{Butler11}
Butler, A.~H. and L.~M. Polvani, 2011: {El Ni\~{n}o, La Ni\~{n}a, and
  stratospheric sudden warmings: A reevaluation in light of the observational
  record}. \textit{Geophysical Research Letters}, \textbf{38~(13)},
  \doi{10.1029/2011GL048084},
  \urlprefix\url{http://doi.wiley.com/10.1029/2011GL048084}.

\bibitem[{Chen and Tung(2014)}]{Chen14}
Chen, X. and K.-K. Tung, 2014: {Varying planetary heat sink led to
  global-warming slowdown and acceleration}. \textit{Science},
  \textbf{345~(2013)}, 897--903, \doi{10.1126/science.1254937},
  \urlprefix\url{http://www.sciencemag.org/cgi/doi/10.1126/science.1254937}.

\bibitem[{Clette et~al.(2014)Clette, Svalgaard, Vaquero, and Cliver}]{Clette14}
Clette, F., L.~Svalgaard, J.~M. Vaquero, and E.~W. Cliver, 2014: {Revisiting
  the Sunspot Number}. \textit{Space Science Reviews}, 35--103,
  \doi{10.1007/s11214-014-0074-2}, \eprint{1407.3231}.

\bibitem[{Collimore et~al.(2003)Collimore, Martin, Hitchman, Huesmann, and
  Waliser}]{Collimore03}
Collimore, C.~C., D.~W. Martin, M.~H. Hitchman, A.~Huesmann, and D.~E. Waliser,
  2003: {On the relationship between the QBO and tropical deep convection}.
  \textit{Journal of Climate}, \textbf{16}, 2552--2568,
  \doi{10.1175/1520-0442(2003)016<2552:OTRBTQ>2.0.CO;2}.

\bibitem[{Crooks and Gray(2005)}]{Crooks05}
Crooks, S.~A. and L.~J. Gray, 2005: {Characterization of the 11-year solar
  signal using a multiple regression analysis of the ERA-40 dataset}.
  \textit{Journal of Climate}, \textbf{18}, 996--1015,
  \doi{10.1175/JCLI-3308.1}.

\bibitem[{Czaja and Frankignoul(2002)}]{Czaja02}
Czaja, A. and C.~Frankignoul, 2002: {Observed impact of Atlantic SST anomalies
  on the North Atlantic oscillation}. \textit{Journal of Climate}, \textbf{15},
  606--623, \doi{10.1175/1520-0442(2002)015<0606:OIOASA>2.0.CO;2}.

\bibitem[{{Dameris} and {Ebel}(1990)}]{Dameris90}
{Dameris}, M. and A.~{Ebel}, 1990: {The quasi-biennial oscillation and major
  stratospheric warmings - A three-dimensional model study}. \textit{Annales
  Geophysicae}, \textbf{8}, 79--85.

\bibitem[{Delworth and Knutson(2000)}]{Delworth00}
Delworth, T.~L. and T.~R. Knutson, 2000: {Simulation of Early 20th Century
  Global Warming}. \textit{Science}, \textbf{287~(March)}, 2246--2250,
  \doi{10.1126/science.287.5461.2246}.

\bibitem[{Elsner et~al.(2010)Elsner, Jagger, and Hodges}]{Elsner10}
Elsner, J.~B., T.~H. Jagger, and R.~E. Hodges, 2010: {Daily tropical cyclone
  intensity response to solar ultraviolet radiation}. \textit{Geophysical
  Research Letters}, \textbf{37}, \doi{10.1029/2010GL043091}.

\bibitem[{Elsner et~al.(1999)Elsner, Kara, and Owens}]{Elsner99}
Elsner, J.~B., a.~B. Kara, and M.~a. Owens, 1999: {Fluctuations in North
  Atlantic hurricane frequency}. \textit{Journal of Climate},
  \textbf{12~(Tinsley 1988)}, 427--437,
  \doi{10.1175/1520-0442(1999)012<0427:FINAHF>2.0.CO;2}.

\bibitem[{Fereday et~al.(2012)Fereday, Maidens, Arribas, Scaife, and
  Knight}]{Fereday12}
Fereday, D.~R., A.~Maidens, A.~Arribas, a.~a. Scaife, and J.~R. Knight, 2012:
  {Seasonal forecasts of northern hemisphere winter 2009/10}.
  \textit{Environmental Research Letters}, \textbf{7~(3)}, 034\,031,
  \doi{10.1088/1748-9326/7/3/034031},
  \urlprefix\url{http://stacks.iop.org/1748-9326/7/i=3/a=034031?key=crossref.65c78d27344f958979668715e18d4353}.

\bibitem[{Free and Seidel(2009)}]{Free09}
Free, M. and D.~J. Seidel, 2009: {Observed El Ni\~{n}o–Southern Oscillation
  temperature signal in the stratosphere}. \textit{Journal of Geophysical
  Research}, \textbf{114~(D23)}, D23\,108, \doi{10.1029/2009JD012420},
  \urlprefix\url{http://doi.wiley.com/10.1029/2009JD012420}.

\bibitem[{Gastineau et~al.(2013)Gastineau, D'Andrea, and
  Frankignoul}]{Gastineau13}
Gastineau, G., F.~D'Andrea, and C.~Frankignoul, 2013: {Atmospheric response to
  the North Atlantic Ocean variability on seasonal to decadal time scales}.
  \textit{Climate Dynamics}, \textbf{40}, 2311--2330,
  \doi{10.1007/s00382-012-1333-0}.

\bibitem[{Hitchcock et~al.(2013)Hitchcock, Shepherd, and Manney}]{Hitchcock13}
Hitchcock, P., T.~G. Shepherd, and G.~L. Manney, 2013: {Statistical
  characterization of Arctic polar-night jet oscillation events}.
  \textit{Journal of Climate}, \textbf{26}, 2096--2116,
  \doi{10.1175/JCLI-D-12-00202.1}.

\bibitem[{Ho et~al.(2009)Ho, Kim, Jeong, and Son}]{Ho09}
Ho, C.~I., H.~S. Kim, J.~H. Jeong, and S.~W. Son, 2009: {Influence of
  stratospheric quasi-biennial oscillation on tropical cyclone tracks in the
  western North Pacific}. \textit{Geophysical Research Letters},
  \textbf{36~(January)}, 2--5, \doi{10.1029/2009GL037163}.

\bibitem[{Hodges et~al.(2014)Hodges, Jagger, and Elsner}]{Hodges14}
Hodges, R.~E., T.~H. Jagger, and J.~B. Elsner, 2014: {The sun-hurricane
  connection: Diagnosing the solar impacts on hurricane frequency over the
  North Atlantic basin using a space-time model}. \textit{Natural Hazards},
  \textbf{73}, 1063--1084, \doi{10.1007/s11069-014-1120-9}.

\bibitem[{Hoerling et~al.(2001)Hoerling, Hurrell, and Xu}]{Hoerling01}
Hoerling, M.~P., J.~W. Hurrell, and T.~Xu, 2001: {Tropical origins for recent
  North Atlantic climate change.} \textit{Science (New York, N.Y.)},
  \textbf{292~(April)}, 90--92, \doi{10.1126/science.1058582}.

\bibitem[{Huesmann and Hitchman(2001)}]{Huesmann01}
Huesmann, A.~S. and M.~H. Hitchman, 2001: {The stratospheric quasi-biennial
  oscillation in the NCEP reanalyses: Climatological structures}.
  \textit{Journal of Geophysical Research}, \textbf{106}, 11\,859--11\,874.

\bibitem[{Kodera and Kuroda(2002)}]{Kodera02}
Kodera, K. and Y.~Kuroda, 2002: {Dynamical response to the solar cycle}.
  \textit{Journal of Geophysical Research}, \textbf{107~(D24)},
  \doi{10.1029/2002JD002224},
  \urlprefix\url{http://doi.wiley.com/10.1029/2002JD002224}.

\bibitem[{Marshall and Scaife(2009)}]{Marshall09}
Marshall, A.~G. and A.~A. Scaife, 2009: {Impact of the QBO on surface winter
  climate}. \textit{Journal of Geophysical Research}, \textbf{114~(D18)},
  D18\,110, \doi{10.1029/2009JD011737},
  \urlprefix\url{http://doi.wiley.com/10.1029/2009JD011737}.

\bibitem[{McIntyre and Palmer(1983)}]{McIntyre83}
McIntyre, M.~E. and T.~N. Palmer, 1983: {Breaking planetary waves in the
  stratosphere}. \textit{Nature}, \textbf{305~(5935)}, 593--600,
  \doi{10.1038/305593a0},
  \urlprefix\url{http://www.nature.com/doifinder/10.1038/305593a0}.

\bibitem[{Pascoe et~al.(2005)Pascoe, Gray, Crooks, Juckes, and
  Baldwin}]{Pascoe05}
Pascoe, C.~L., L.~J. Gray, S.~a. Crooks, M.~N. Juckes, and M.~P. Baldwin, 2005:
  {The quasi-biennial oscillation: Analysis using ERA-40 data}. \textit{Journal
  of Geophysical Research D: Atmospheres}, \textbf{110~(December 2004)},
  \doi{10.1029/2004JD004941}.

\bibitem[{Qun and Qiuming(1993)}]{Qun93}
Qun, X. and Y.~Qiuming, 1993: {Response of the intensity of subtropical high in
  the northern hemisphere to solar activity}. \textit{Advances in Atmospheric
  Sciences}, \textbf{10~(3)}, 325--334, \doi{10.1007/BF02658138}.

\bibitem[{Rhein et~al.(2013)}]{IPCC13}
Rhein, M., et~al., 2013: {Climate Change 2013: The Physical Science Basis,
  Chapter 3 Observations: Ocean}. \textit{Intergovernmental Panel on Climate
  Change}, 255--315, \doi{10.1017/CBO9781107415324.010}.

\bibitem[{Rind et~al.(2005)Rind, Perlwitz, and Lonergan}]{Rind05}
Rind, D., J.~Perlwitz, and P.~Lonergan, 2005: {AO/NAO response to climate
  change: 1. Respective influences of stratospheric and tropospheric climate
  changes}. \textit{Journal of Geophysical Research D: Atmospheres},
  \textbf{110}, 1--15, \doi{10.1029/2004JD005103}.

\bibitem[{Rodwell(2003)}]{Rodwell03}
Rodwell, M.~J., 2003: {On the predictability of the North Atlantic ocean
  state}. \textit{Geophysical Monograph}, \textbf{134}, 173--192.

\bibitem[{Scaife et~al.(2013)Scaife, Ineson, Knight, Gray, Kodera, and
  Smith}]{Scaife13}
Scaife, A.~a., S.~Ineson, J.~R. Knight, L.~Gray, K.~Kodera, and D.~M. Smith,
  2013: {A mechanism for lagged North Atlantic climate response to solar
  variability}. \textit{Geophysical Research Letters}, \textbf{40}, 434--439,
  \doi{10.1002/grl.50099}.

\bibitem[{Soukharev and Hood(2001)}]{Soukharev01}
Soukharev, B.~E. and L.~L. Hood, 2001: {Possible solar modulation of the
  equatorial quasi-biennial oscillation: Additional statistical evidence}.
  \textit{Journal of Geophysical Research}, \textbf{106}, 14,855--14,868.

\bibitem[{White et~al.(1997)White, Lean, Cayan, and Dettinger}]{White97}
White, W.~B., J.~Lean, D.~R. Cayan, and M.~D. Dettinger, 1997: {Response of
  global upper ocean temperature to changing solar irradiance}. \textit{Journal
  of Geophysical Research}, \textbf{102}, 3255--3266.

\end{thebibliography}

\end{document}